# Real-time imaging of standing-wave patterns in microresonators

Haochen Yan[1,2], Alekhya Ghosh[1,2], Arghadeep Pal[1,2], Hao Zhang[1], Toby Bi[1,2], George Ghalanos[1], Shuangyou Zhang[1], Lewis Hill[1,3], Yaojing Zhang[1], Yongyong Zhuang[1,4], Jolly Xavier[1,5], Pascal Del'Haye[1,2]*

[1]*Max Planck Institute for the Science of Light, Staudtstr. 2, 91058, Erlangen, Germany*
[2]*Department of Physics, Friedrich Alexander University Erlangen-Nuremberg, 91058, Germany*
[3]*SUPA and Department of Physics, University of Strathclyde, 107 Rottenrow, Glasgow G4 0NG, United Kingdom*
[4]*Electronic Materials Research Laboratory, Key Laboratory of the Ministry of Education & International Center for Dielectric Research, School of Electronic Science and Engineering, Faculty of Electronic and Information Engineering, Xi'an Jiaotong University, Xi'an 710049, China*
[5]*SeNSE, Indian Institute of Technology Delhi, Hauz Khas, New Delhi, India*

*Corresponding author: Pascal Del'Haye

**Email:** pascal.delhaye@mpl.mpg.de

**Abstract**
Real-time characterization of microresonator dynamics is important for many applications. In particular it is critical for near-field sensing and understanding light-matter interactions. Here, we report camera-facilitated imaging and analysis of standing wave patterns in optical ring resonators. The standing wave pattern is generated through bi-directional pumping of a microresonator and the scattered light from the microresonator is collected by a short-wave infrared (SWIR) camera. The recorded scattering patterns are wavelength dependent, and the scattered intensity exhibits a linear relation with the circulating power within the microresonator. By modulating the relative phase between the two pump waves, we can control the generated standing waves' movements and characterize the resonator with the SWIR camera. The visualized standing wave enables subwavelength distance measurements of scattering targets with nanometer-level accuracy. This work opens new avenues for applications in on-chip near-field (bio-)sensing, real time characterization of photonic integrated circuits and backscattering control in telecom systems.

**Main Text**
**Introduction**
Whispering-gallery-mode (WGM) microresonators[1,2] are versatile platforms for studying nonlinear optical physics due to their small mode volumes and ultrahigh quality factors. They have been responsible for recent advances in applications such as comb generation[3-8], on-chip lasers[9-11], biomedical sensing[12-18], and optical communications[19-21]. In a lot of applications, backscattering and the resulting generation of standing waves generated inside a microresonator is a topic of great interest[22-27]. It can not only promote our understanding of fundamental physics, such as strong coupling between the atom and microcavity[28], symmetry breaking phenomena[29-33] and cavity quantum electrodynamics (cavity QED)[34,35], but also be a fascinating tool to realize various applications[36,37]. For instance, standing waves can be used for precision sensing, in which slight variations of the forward and backward propagating laser fields can significantly alter the distribution of the interference pattern inside the cavity[12,14,15]. In addition, in recent years there has been growing interest in live visualization of microcavity because of the substantial dynamical information provided about the system[38]. Specifically, applying imaging methods to investigate other phenomena such as field distribution[39] and second-order harmonic generation[40] have been reported. Another example is the use of slow-light[41-43] for imaging, with potential applications in optical fiber fabrication. Thus, the real-time imaging of scattering patterns is urgently demanded since it could enable novel ways of learning intra-cavity dynamics[44,45] such as characterizing dissipative Kerr soliton states and soliton crystals states in microresonators.

In this article, we present a novel approach involving a camera-facilitated visualization technique to investigate the generation and manipulation of the standing wave in toroidal WGM microresonators. In a bi-directional pumping scheme, a short-wave infrared (SWIR) camera is utilized to collect scattered light, which we use to characterize the standing waves in the microresonator. This live-visualization approach enables precise studies of the temporal evolution of the standing wave patterns. In addition, subwavelength distance measurements are conducted by moving the standing wave maxima along the circumference of the resonator and measuring the visibility of scatterers at different locations. This enables nanometer-level localization of scatterers. To our knowledge, this is the first demonstration of real-time characterization of standing wave patterns utilizing a SWIR camera. This measurement technique could be useful for bio-sensing, resonator characterization and real-time monitoring of soliton states.

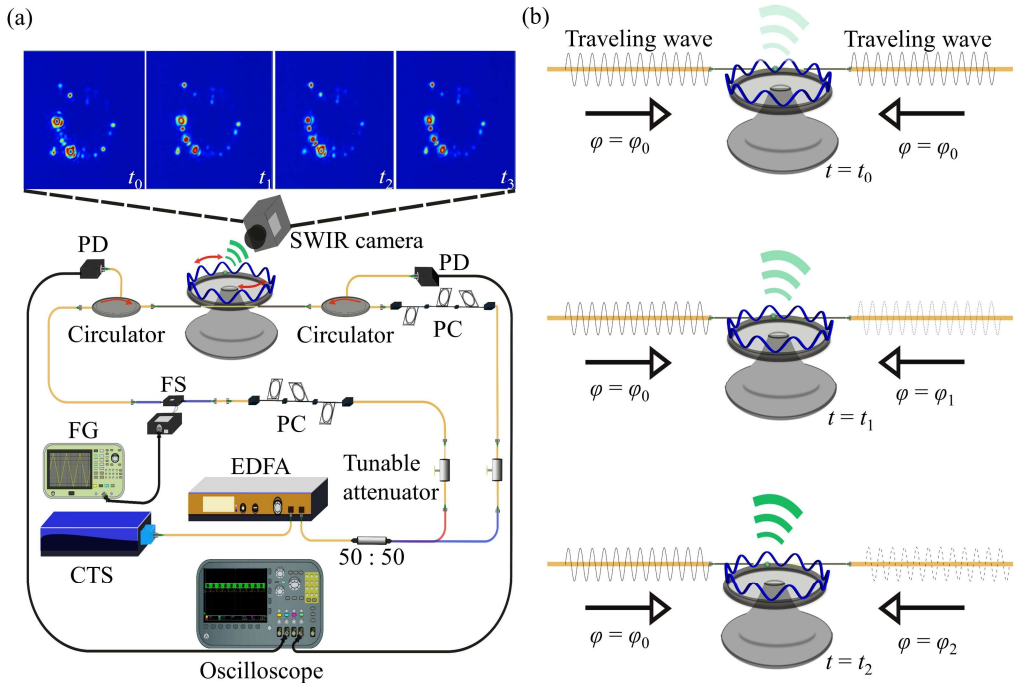

**Figure 1**. Experimental setup and principle. (a) Scheme of the experimental setup to generate and observe standing wave scattering. SWIR: short-wave infrared; PD: photodiode; PC: polarization controller; FS: fiber stretcher; FG: Function generator CTS: continuously-tunable laser. (b) Principle of standing wave generation and characterization. For a specific scatterer, the scattered light intensity is maximized when its physical position matches the standing wave crest, while the intensity is minimized when its physical position matches the standing wave trough. The standing wave is manipulated by changing the relative phase between the bi-directional pump waves so that the wave crest and trough can move across the specific scatterer.

**Results**

**Experimental setup and principle**

The experimental setup and standing-wave generation principle are schematically depicted in Fig. 1(a). A continuously tunable laser (CTS) with a wavelength of 1.5 μm is used as the pumping source to generate the standing waves. The laser light is first amplified by an Erbium-doped fiber amplifier (EDFA) and then divided by a 50/50 coupler in order to generate two light waves that are used for counterpropagating coupling into a microresonator. Both directions' polarization states and power levels are controlled independently with tunable attenuators and polarization controllers (PC). A 130-μm-diameter fused silica toroidal microresonator (Q factor ~$10^6$) is coupled to the system with a tapered fiber coupler[46]. The position of the tapered fiber is finely adjusted by a XYZ piezo stage until critical coupling is achieved[2]. A fiber stretcher is connected to one of the pumping paths and modulated by a function generator to alter the phase difference between the two paths. The relative phase between the counterpropagating light waves is used to precisely control the position of the maxima of the standing wave within the resonator. A SWIR camera (NIT SenS 1280) images the light scattered out from the microresonator. The generated standing wave can be characterized and manipulated by tuning the relative phase between the two-direction pumping waves. The characterization principle is illustrated in Fig. 1(b). The measurement with the SWIR camera depends on the overlap between a scatterer's physical position with the maximum of a standing wave. If the scatterer is close to a minimum of the standing wave, only very little scattered light can be detected.

**Characterization of scattered intensity**

In the experiments, we first characterize the recorded scattering intensities from the SWIR camera (50 μs exposure time with 50 Hz frame rate, same for all the other results). In particular, we study the scattered intensities recorded at different coupling conditions and different laser wavelengths. The detailed experimental procedure is explained in Methods, and the selected images of different coupling and pumping conditions are shown in Fig. 2(a). We couple light into the resonator bi-directionally at different wavelengths and different physical coupling position, which results in changing scattering patterns. We observe that for a similar pumping wavelength, the scattering patterns share some similarities even if the physical coupling position is distinct (i.e., the taper couples from left-hand side or the taper couples from right-hand side), as shown by the first and second image in Fig. 2(a), where the strongest scatter intensity appears at a similar position. On the contrary, the scattering patterns are quite distinct when the pumping wavelength is different even if the physical coupling position remained the same, as shown by the first and third image in Fig. 2(a). This indicates that the scattering pattern is mostly pump-wavelength dependent, which agrees with the detection principle depicted in Fig. 1(b). The statistical information of the scattered intensity, as a function of the angle, is shown in Fig. 2(b). The differences of scattered intensity between these two points are also plotted and shown in Fig. 2(b). Saturations are observed for specific positions, and these points are avoided in further analysis due to their artifacts.

We further study the relationship between scattered light intensity and the power coupled into the cavity (see Methods). Specifically, we set the laser pumping wavelength to 1565 nm, and the laser wavelength is then finely tuned into the resonance of the microtoroid from the blue-detuned to the red-detuned side. We stop at several specific detuning positions and record the transmitted power collected from the photodiodes (PDs) as well as the scattered light intensity from the SWIR camera. A recorded video with continuously changing detuning can be found in Supplementary Video 1. The recorded transmitted power and scattering intensities for regions of interest are measured as functions of detuning and are shown in Fig. 2(c), where different colors stand for different selected regions. To further confirm the linearity, we plot the scattered intensities for all selected regions as the function of coupled power, as shown in Fig. 2(d). The direct linear relation between coupled power and scatted intensity presented in this graph solidifies our conclusion above.

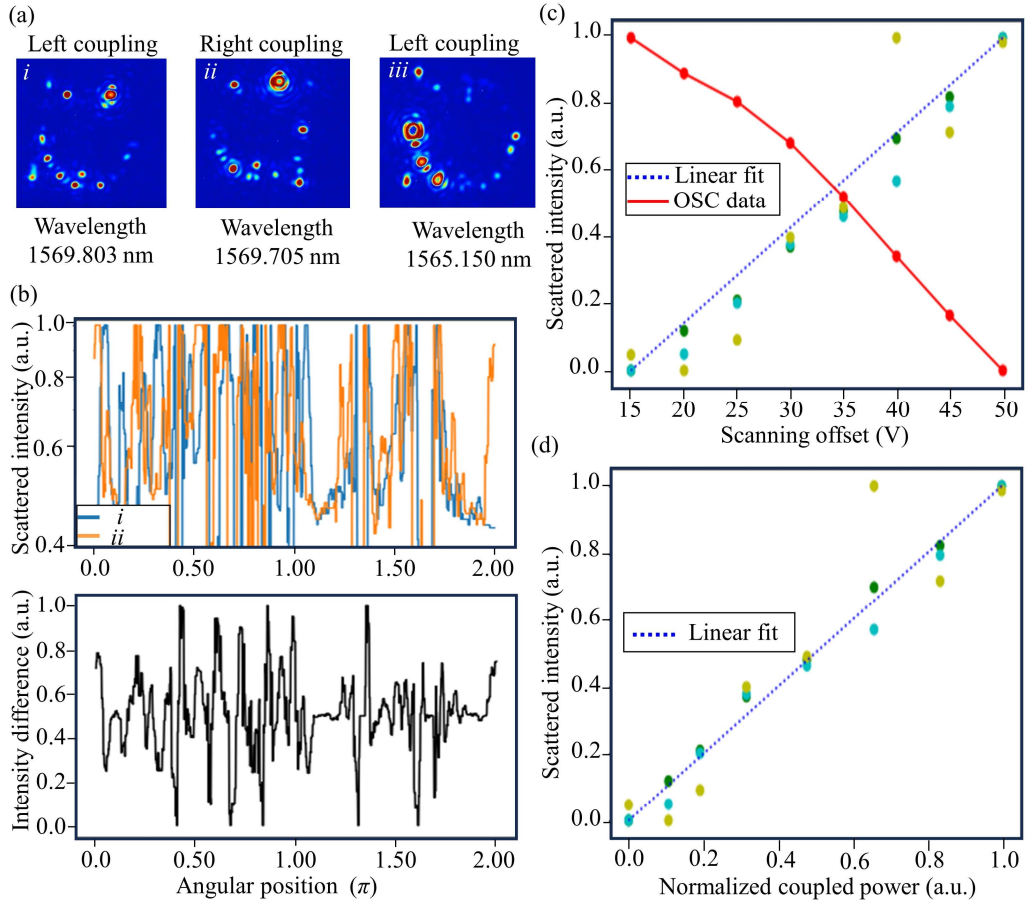

**Figure 2.** Characterization of the scattering from a microresonator. (a) Images of the SWIR camera at different coupling conditions and pump wavelengths. (b) Scattering intensity as a function of angle around the resonator based on the first two images in panel (a). The optical intensity for each angular position is plotted. The difference of scattered intensity between image i and image ii is shown in the lower part of panel (b). Note that the camera is saturated in some positions with strong scattering. (c) Transmitted power and scattering intensity of different points as a function of laser detuning. (d) Scattered power as a function of coupled power. We observe a linear increase of the scattered light with increasing intracavity power.

**Simulation of the standing wave scattering**

To theoretically investigate the standing wave generation and the scattering of the standing wave induced by defects, we perform a finite element method (FEM) simulation (COMSOL) of the bi-directionally pumped microresonator. The electric field distribution (E field norm, automatically normalized by COMSOL) across the microresonator, generated from bi-directional pumping (wavelength 1566.5 nm) in a 2D microresonator structure, is shown in Fig. 3(a). We observe discrete mode distribution in the microresonator, which corresponds to the standing wave. Moreover, we designed a defect with a circular shape on the core material of the structure to mimic a scatterer on the microresonator, and by varying its radius (700 nm and 800 nm), we obtained different mode distributions, as shown in Fig. 3(a) as well, where the position of the scatterer is marked with a red rectangular box. Considering that the SWIR camera resolution is not high enough to identify each scatterer independently, and there may be several scatterers located closely within a small region, the simulation results here can be reasonably matched to what we observe from the SWIR camera. Finally, we vary the scatterer position and check the mode distributions when the scatterer size is fixed (800 nm), as shown in Fig. 3(b). We start with a specific position (panel one) and then move the scatterer's position by 3/10 of the mode's wavelength and 6/10 of the mode's wavelength respectively (panel two and panel three). From these simulations we can see that the position of the scatterer influences the mode pattern the resonator.

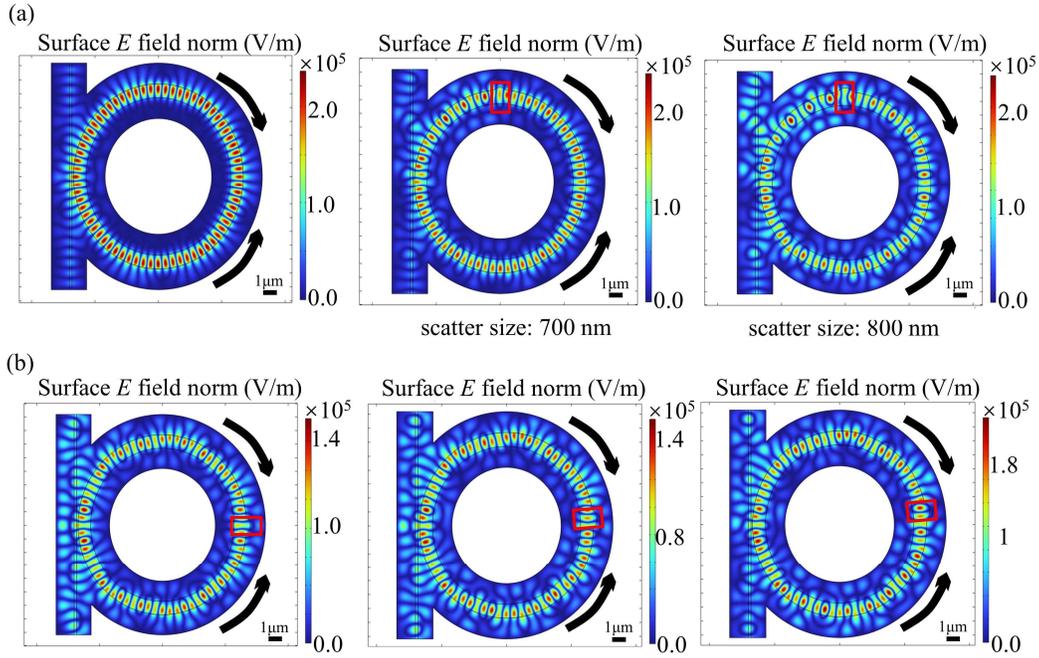

**Figure 3.** Simulation of dual pumped toroidal microresonators. (a) Bi-directional pumping pattern with and without defects. Defects are designed as a half circle with different diameters: 700 nm and 800 nm. (b) Standing wave patterns for different defect positions with one single defect with a size of 800 nm. The defect position in the left panel is considered as reference, while the defect position is upshifted 1/10 λ (resonance wavelength) and 6/10 λ in the middle and right panels, respectively.

**Moving the nodes of the standing wave**
We perform a detailed study to demonstrate and characterize this rotation of standing wave distribution in the microresonator with the mechanism stated in Fig.1(b). Specifically, we modify the amplitude and frequency of the triangle signal that modulates the relative phase relation between the two pump waves. As discussed earlier, two counterpropagating beams with the same frequency form a standing wave, which is characterized by the positions of nodes and anti-nodes of the standing wave pattern. Changing the relative phase of one of the beams effectively changes the position of the nodes and anti-nodes of the distribution. During the modulation period, the scattered light intensity is recorded as a video, and we perform a fast Fourier transform (FFT) of the recorded data to obtain the standing wave oscillation frequencies. The FFT results of recorded data with two modulation frequencies 3 Hz and 5 Hz are shown in Fig. 4(a) and 4(b), respectively. Three modulation amplitudes (peak-to-peak voltage) are used for each modulation frequency, and we find that the scattered light oscillates with the exact frequency of phase modulation. This can be attributed to the fact that a change of phase causes the standing wave to rotate along the periphery of the resonator, and thus causes sinusoidal fluctuation of the light scattered from the scatterers who present inside the optical mode. For these additional frequency peaks, there are two major causes. Firstly, at the end of each period where the ramp dives the phase shifter, there is an abrupt jump in the voltage, which arises a sudden discontinuity of modulation. Correspondingly, the relative phase and thus the scattered light intensity change nonlinearly for each period, which orders higher harmonics in the frequency domain for the FFT. Moreover, the scattered light from the scatterers completes a period of sinusoid when each time the maxima of the standing wave in the resonator moves across it. For each ramp modulation of the phase shifter, many periods of scattered light sinusoids are completed. Thus, the frequency components present integer multiples of the ramp frequency. Since the phase modulation moves the standing wave at a predefined frequency, we confirm that this movement is imprinted onto the magnitude of scattered light, which depends on the exact position of the scatterers with respect to the standing wave maxima. The recorded videos for 3 Hz and 5 Hz modulation can be found as Supplementary Videos 2 and 3 respectively.

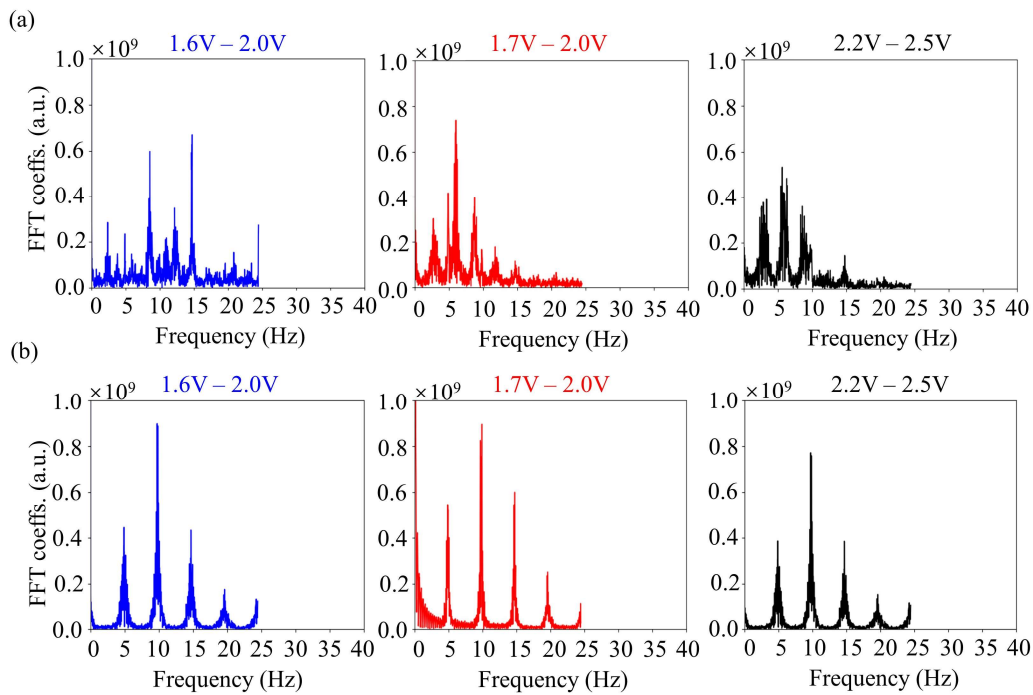

**Figure 4.** Characterization of standing waves. (a) FFT of the scattered light intensity from the video with 3 Hz modulation with three different modulation amplitudes. (b) FFT of the scattered light intensity from the video with 5 Hz modulation and three different modulation amplitudes. All the modulation amplitudes refer to the peak-to-peak voltage of the applied triangle signal.

**Correlation studies and applications**

SWIR camera based standing wave visualization can be applied in various areas, especially for investigating cavity-enhanced light-matter interactions. Here we demonstrate a simple scheme of utilizing the standing wave, as shown in Fig. 5(a). The intrinsic phase relation between two scatterers can be visualized and quantitatively studied by correlating the phase of the standing wave with the scattered light intensities. In the experiment, we record the scattered light intensity oscillations over several cycles of phase modulation (5Hz modulation frequency) and analyze the phase offset for selected regions of interest. Two snapshot images are presented in Fig. 5(b), where the regions of interest are marked on the images. For each image, a few specific regions (marked as 1,2,3 and i,ii,iii,iv respectively) with the same pixel dimensions are selected for further quantitative study of the phase relation between different scattering regions. Notably, we plot the intensities of different two points as a function of time in phase space (see Methods), as shown in Fig. 5(c). We observe three well-shaped ellipses which represent the in-phase (red and green) and out-phase (yellow) relation respectively. In addition, a nearly perfect in-phase relation is also picked and demonstrated in the panel with blue points, where a more defined linear shape rather than an elliptical shape is observed. Note, a completely in-phase (out-of-phase) oscillation would result in a straight line passing through the origin with slope $\pi/4$ ($3\pi/4$) in phase space. Oscillations with the same amplitude but constant phase difference yield an ellipse, with either the major or minor axis along the straight line passing through the origin with slope π/4. The measured standing waves scattering intensity can be used for precise distance measurement between the scatterers. More specifically, we show how to use the fit of the obtained phase-space ellipses to extract the relative phase difference and then calculate the distance between these two points (see Methods). We show this with the selected two pairs of points marked in the left panels in Fig. 5(d). The corresponding phase-space plot and fits are shown in the right panels of Fig. 5(d). The measured distances of the scatterers based on the images are 223.80 μm (A and B) and 65.73 μm (C and D), respectively, where their corresponding sub-wavelength distances based on the phase of the emitted intensities are 189.4 nm and 193.9 nm (see Methods). The calculated distances' uncertainties are as small as 5 nm and 5.02 nm, which corresponds to ~1/300th of the pump wavelength. We also tested the repeatability of the distance measurements over the course of one day. After one day, the calculated coarse distances are the same, but the fine distances show a difference of 15.6 nm. This difference can be attributed to the following factors: 1) A change in the intrinsic phase relation fitting between points due to the slight change of the coupling conditions; 2) A shift of the captured points' positions in the image that leads to a change in signal to noise in the measurement of the scattered power; 3) A physical deformation of the microresonator itself due to lab temperature changes; and 4) A shift of the scatterer's physical position due to a slow change of the silica's amorphous structure. Our results demonstrate a simple and elegant method for real-time characterization of standing wave patterns, which can be quantitively studied for highly-accurate distance measurements.

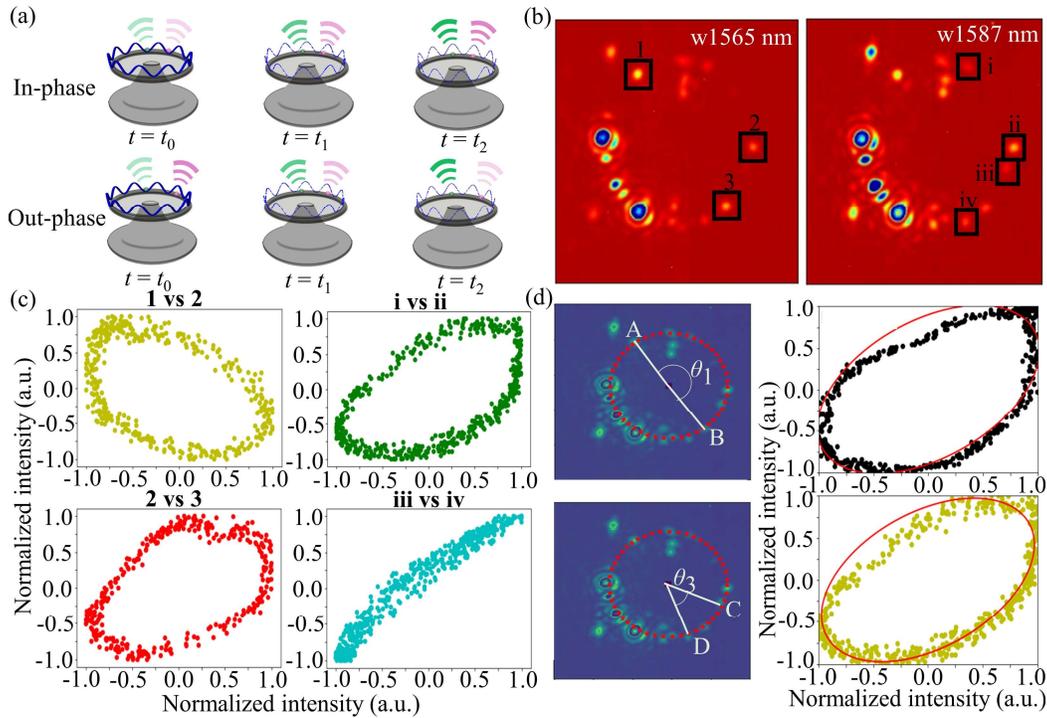

**Figure 5.** Characterization of the correlation between different scattering points. (a) Scheme of applying standing waves for distance measurements. The scatterers' distance can be inferred from the relative phase of the scattered light intensity; (b) Selected snapshot images with labeled scatterers. The selected regions do not show saturation during the data recording. (c) Phase space plots of four sets of selected pairs of points. In-phase and out-of-phase relations between scattered light powers can be seen between different points. (d) Sub-wavelength accuracy distance calculation for selected points. The precise distances between scatterers are calculated from the fitting parameters of the ellipses in the phase space plots.

**Discussion**

In conclusion, we demonstrate real-time analysis of standing wave patterns in microresonators. With phase control of the input light waves we can control the position of the standing wave maxima to trace out scatterers within a microresonator. The visualization of the scattered light from the standing wave can be utilized to perform distance measurements of scatterers with subwavelength accuracy. In our proof of principle measurements, we determine the relative position between different scatterers with a precision of ~5 nm. This work enables super-resolution characterization and quantization of intracavity dynamics. The real-time characterization of optical microresonators with standing waves can be used for sensing and other advanced integrated nanophotonic applications. For instance, on-purpose fabricated defects (e.g., nano-antennas) on the microresonator can be precisely and selectively activated by manipulating the standing wave[47]. This can also be applied to quantum emitters like quantum dots within the microresonator[48-50]. Coherent optical activation of a large number of emitters on a microresonator could enable new types of coherent light sources. Moreover, applying the standing wave for subwavelength measurements is expected to advance microresonator-based on-chip sensing[15,51,52] by providing precise spatial information that cannot be accessed by either transmission or reflected signals currently utilized. In addition to designed defects, the current study can be extended further to understand binding events of molecules to the resonator surface as well as transient light-matter interactions[53]. Finally, the method proposed in our study can be considered a handy tool for real-time characterization of scattering sources in integrated photonic chips, which can benefit industrial applications.

## Material and Methods
### Experimental procedure

For these experiments, we first optimize the coupling position by subtly adjusting the XYZ piezo-controlled stage as well as the pumping wavelength to achieve critical coupling conditions. Then the polarization states are independently optimized for the two pumping directions. Once the coupling is optimized, we detune the laser by continuously decreasing its scanning range to zero and subtly adjusting the pumping wavelength. Finally, the power coupled to the cavity is maximized, i.e., the transmitted power observed on the oscilloscope is minimized. For all experiments, laser scanning is turned off. For the scattering intensity characterization experiment, measurements are conducted at different detuning steps, while for all the other cases, the measurements are conducted when the coupling power is maximized. The camera (either visible or 1550-nm) is placed vertically above the sample, and the imaging system consists of a zoom lens (Thorlabs, 6.5X with 0.006 – 0.142 system NA) an extension tube (Thorlabs, 2.0X magnification) and a magnification lens (Thorlabs, 2.0X magnification) to obtain sufficient magnification to see the microresonator. The imaging system is installed on a 3-D translation stage. We first use the visible-light CMOS camera (Thorlabs, CS165CU/M) to identify the position of the microresoantor. Then we replace the visible-light camera with the 1550-nm camera and slightly tune the camera height (i.e., z/vertical direction) to refocus on the microresonator at the new wavelength. For the standing wave characterization experiments, we take videos for different modulation conditions and trace one specific region's intensity change during a certain time interval. Then we perform FFT of the collected data to obtain the frequency results. For the data analysis, we use grayscale raw images. The raw images saved by the camera have a linear scale (ranging from 0 to 16,383 corresponding to 14 bits). In the images shown in the main text we applied a standard Jet colormap to the data, in order to make it easier to identify the scattering patterns.

### Theoretical simulation model

To observe the optical field within the microresonator, we conduct the simulation by using the finite element method (FEM) via the COMSOL software. To reduce the simulation time, we chose a reasonably small dimension of the resonator. Specifically, the structure is considered two-dimensional, and the radius and width of the core of the resonator are chosen as 6.2 µm and 1 µm respectively. The core material is the same as the experimental case, i.e., fused silica with no cladding material. The light enters the microresonator owed to the bidirectional pumping from the tapered fiber, which itself has a width of around 200 nm. The distance between the center of the resonator's core and the taper's center is selected to be around 716 nm, which shows a considerable coupling of light into the system. The scatterer is introduced in the system in terms of having a structural variation or defect in the core.

### Phase space plots and distance calculation

A rotating standing wave induced by the changing phase in one of the arms will enforce a phase difference between two scatterers positioned at two different points on the microresonator periphery. To study this effect, we chose a few distinct points on the microresonator demonstrating scattering in the images obtained from the 1550 nm camera. The thermal noise is eliminated by subtracting the average over each 100-time step, and then the data points are normalized to the range -1 to 1. The sinusoidal oscillations of different scatterers with constant phase difference are also observed in the phase space diagrams in Fig. 5c. In each subplot of Fig. 5c, we plot the values of scattered intensities of one scattering point with respect to another scattering point. The equations of the ellipses are derived from two oscillating variables with constant phase difference.

$$x = \alpha \sin \theta$$
$$y = \alpha \sin(\theta + \phi)$$
$$\frac{x^2}{\alpha^2 \sin^2 \phi} - \frac{2xy\sqrt{1-\sin^2 \phi}}{\alpha^2 \sin^2 \phi} + \frac{y^2}{\alpha^2 \sin^2 \phi} = 1.$$

To simplify the calculation and fitting, we consider the ellipse in the polar coordinate such that

$$r^2 \left( \frac{1}{\alpha^2 \sin^2 \phi} - \frac{\sin 2\beta \cos \phi}{\alpha^2 \sin^2 \phi} \right) = 1,$$

and the above equation can be furtherly simplified as

$$r^2 = \frac{\alpha^2 \sin^2 \phi}{1 - \sin 2\beta \cos \phi},$$

where the angle $\phi$ can be directly obtained from fitting the ellipse, while the uncertainty is determined by constructing confidence ellipses (1σ). The physical distance between two points can be calculated from the camera image. By fitting a circle to the collection of scatterers, the pixel distance $l$ between two scatterers on the periphery is given by:

$$l = r \times \Delta\theta$$

Where $\Delta\theta$ is the angle formed by the two points at the center, and r is the radius in terms of pixels. The actual distance between two points, $L$, can be written as,

$$L = \frac{l}{x_p} + \Delta x,$$

where $x_p$ is the length of each pixel, and $\Delta x$ is the sub-pixel distance. This subpixel distance can be measured from the phase difference between the two points, which is given by:

$$\Delta\phi = n\lambda + \phi,$$

where

$$n = \left\lfloor \frac{1}{\lambda} - \frac{l}{x_p} \right\rfloor \quad L - \lambda \left\lfloor \frac{1}{\lambda} - \frac{l}{x_p} \right\rfloor = \frac{\phi}{k},$$

with the standard definition such that $k = 2\pi/\lambda$. The uncertainty of the obtained distance is calculated by propagating the uncertainty of angle $\phi$ obtained from the fitting.

**Acknowledgments**
This work was supported by the European Union's H2020 ERC Starting Grant 756966, the Marie Curie Innovative Training Network "Microcombs" 812818 and the Max Planck Society.